\documentclass[aps,prl,twocolumn,showpacs]{revtex4} 

\usepackage{amsmath,epsfig,amsfonts,epsfig,graphicx,
}
\begin{document}

\title{Defect Modes in One-Dimensional Granular Crystals}

\author{N. Boechler$^{1}$, G. Theocharis$^{1}$, Y. Man$^{1,2}$, P. G. Kevrekidis$^{3}$, and C. Daraio$^{1}$}

\affiliation{ 
$^1$ Graduate Aerospace Laboratories (GALCIT), California Institute of Technology, Pasadena, CA 91125, USA \\
$^2$ Keble College, University of Oxford, Oxford OX1 3PG, UK\\
$^3$ Department of Mathematics and Statistics, University of Massachusetts, Amherst, MA 01003-4515, USA
}


\begin{abstract}
We study the vibrational spectra of one-dimensional statically compressed granular crystals (arrays of elastic particles in contact) containing defects. We focus on the prototypical settings of one or two spherical defects (particles of smaller radii) interspersed in a chain of larger uniform spherical particles. We measure the near-linear frequency spectrum within the spatial vicinity of the defects, and identify the frequencies of the localized defect modes. We compare the experimentally determined frequencies with those obtained by numerical eigen-analysis and by analytical expressions based on few-site considerations. We also present a brief numerical and experimental example of the nonlinear generalization of a single-defect localized mode. 
\end{abstract}



\maketitle



\section{Introduction}
Defect modes in crystals have long been studied in the realm of solid state physics \cite{Maradudin,Lif}. The presence of defects or ``disorder'' is known to enable localized lattice vibrations, whose associated frequencies have been measured in the spectra of real crystals (see \cite{Maradudin,luco} and references therein). More recently, this study has been extended to include other examples, including superconductors \cite{andreev} and electron-phonon interactions \cite{tsironis}. Similar phenomena have also been observed in nonlinear systems, including photonic crystals \cite{Joann}, optical waveguide arrays~\cite{Pes_et99,Mor_et02,kipp}, dielectric superlattices (with embedded defect layers) \cite{soukoulis}, micromechanical cantilever arrays \cite{sievers}, and Bose-Einstein condensates of atomic vapors \cite{engels,anderson}.
 
Granular crystals are nonlinear systems composed of densely-packed particles interacting through Hertzian contacts \cite{Johnson,nesterenko1,sen08,coste97}. These systems present a remarkable ability to tune their dynamic response from linear to strongly nonlinear regimes \cite{nesterenko1}. This has allowed the exploration of fundamental nonlinear waveforms such as traveling waves \cite{nesterenko1,sen08,coste97,our08} and discrete breathers \cite{our10}. Granular crystals have also been proposed for several engineering applications, such as energy absorbing layers \cite{dar06,hong05,fernando,doney06}, actuating devices \cite{dev08}, and sound scramblers \cite{dar05,dar05b}.  

The presence of defects in statically uncompressed (or weakly compressed, as compared to the relative dynamic displacements) granular chains excited by impulsive loading has been studied in a number of previous works that have reported the existence of interesting dynamic responses such as the fragmentation of waves, anomalous reflections, and energy trapping \cite{Job,dar06,hong05,fernando,doney06,Hascoet,hinch99,hong02,sen98,manciu99}. In this paper, we study the response of strongly compressed granular crystals, with one or two defects (extending our earlier theoretical work \cite{Theo2009}), excited by continuous signals. We measure the frequency response of the system and reveal localized modes due to the presence of defects. We report that the number of localized modes mirrors that of the defects, and note that the frequencies of such modes depend on: (i) the ratio of the defect mass to the mass of the particles in the uniform chain, (ii) the relative proximity of multiple defects, (iii) the geometric and material properties of the particles composing the crystal, and (iv) the static load. We compare our experimental findings with numerical computations and with theoretical analysis approximating the behavior of a few sites in the vicinity of the defect(s). Finally, we demonstrate that as we go from the linear to the nonlinear regime, nonlinear ``deformations'' of the linear defect modes (with appropriately downshifted frequencies) are sustained by the system.


\section{Experimental Setup} 
\begin{figure}[h]
\begin{center}
\includegraphics[width=8.3cm,height=2.5cm]{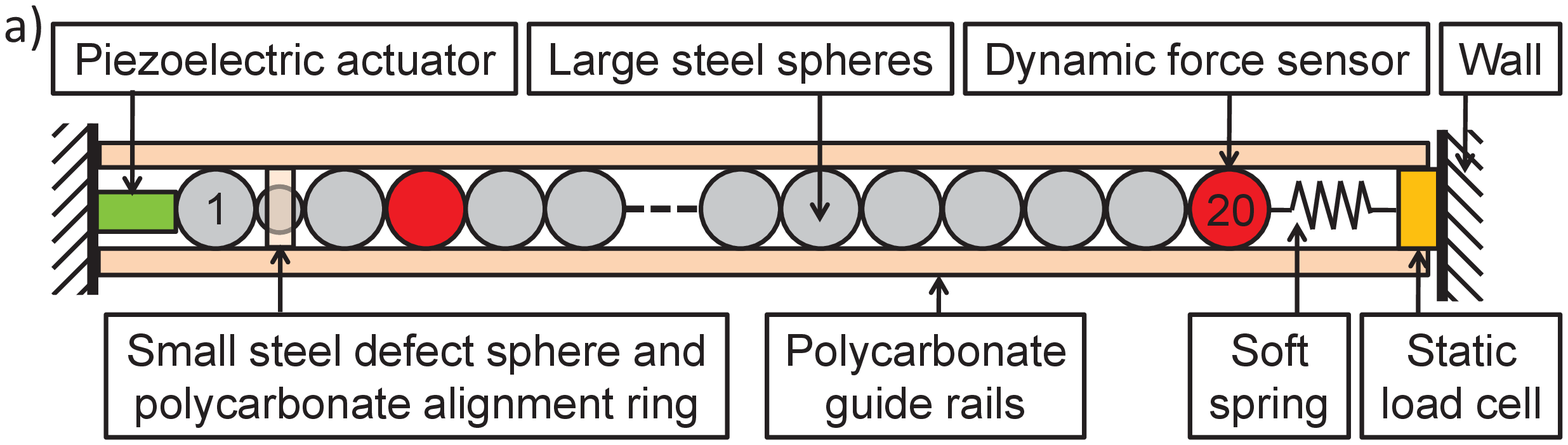}
\includegraphics[width=8.3cm,height=3.5cm]{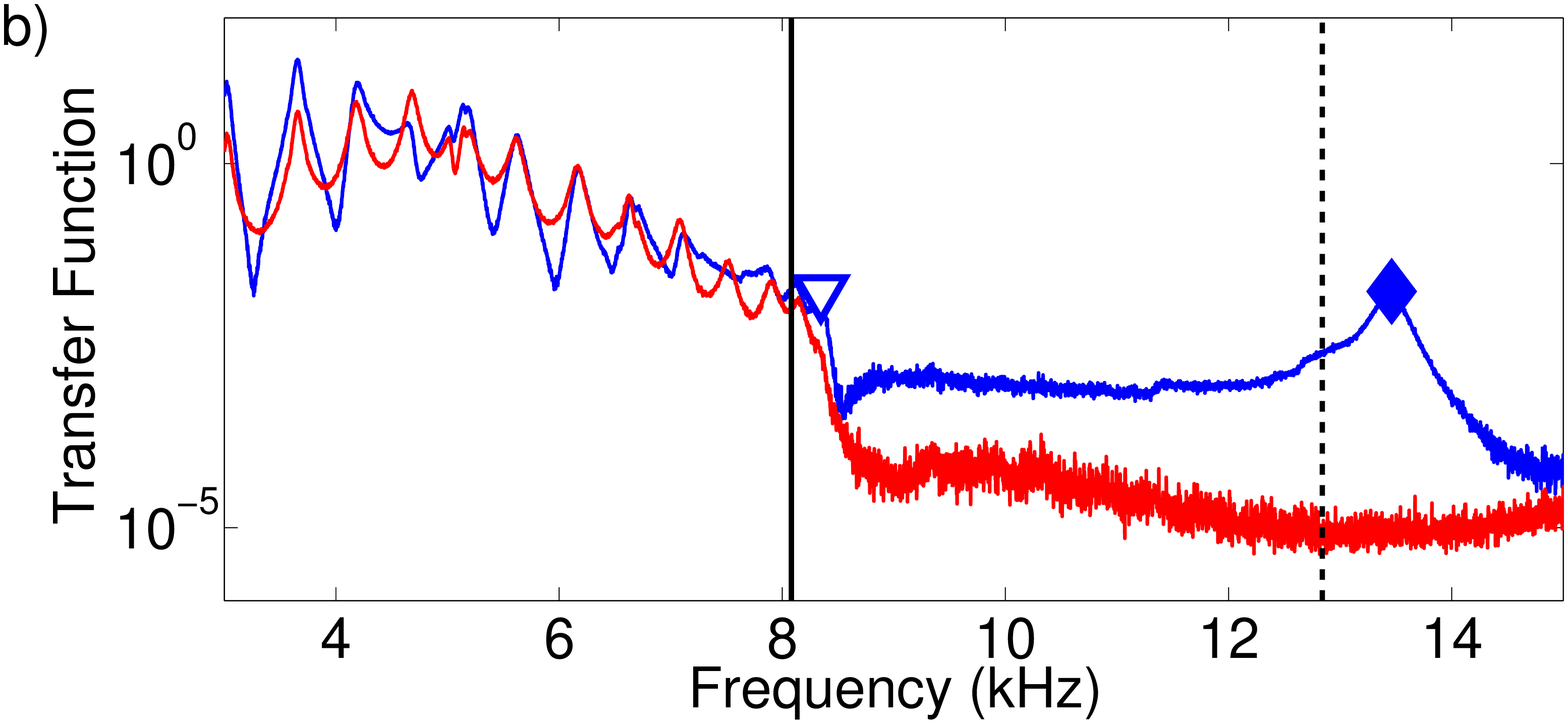}
\end{center}
\caption{\label{setup_spectrum}[Color online] a) Schematic diagram of the experimental setup for the homogeneous chain with a single defect configuration.
b) Experimental transfer functions (as defined in the ``single-defect: near linear regime'' section) for a granular crystal with a static load of $F_0=20$~N and a defect-bead of mass $m=5.73$~g located at site $n_{def}=2$. Blue (dark-grey) [red (light-grey)] curves corresponds to transfer function obtained from the force signal of a sensor particle placed at $n=4$ [$n=20$]. The diamond marker is the defect mode. The triangle marker is the upper acoustic cutoff mode. The vertical black dashed line is the theoretically predicted defect mode frequency, and the vertical solid black line is the theoretically predicted upper acoustic cutoff frequency.}
\end{figure}

We assemble 1D granular crystals, similar to those described in~\cite{our10,our11}, composed of $N=20$ statically compressed stainless steel spherical particles ($316$ type, with elastic modulus $E=193$~GPa and Poisson ratio $\nu_b=0.3$~~\cite{ElasticProperties}), as shown in Fig.~\ref{setup_spectrum}(a). The chain is composed of uniform particles of (measured) radius $R=9.53$~mm and mass $M=28.84$~g, except for one (or two) light-mass stainless steel defect particles. The spheres are held in a 1D configuration using four polycarbonate bars ($12.7$~mm diameter) that are aligned by polycarbonate guide plates spaced at approximately $12$~cm intervals along the axis of the crystal. The defect particles, which are of smaller radii than the rest of the particles of the chain, are aligned with the axis of the crystal using polycarbonate support rings. Dynamic perturbations are applied to the chain by a piezoelectric actuator mounted on a steel cube (which acts as a rigid wall).
The particles are statically compressed by a load of $F_0=20$~N. The static load is applied using a soft spring (of stiffness 1.24 kN/m), which is compressed between the last particle in the chain and a second steel cube bolted to the table. The applied static load is measured by a calibrated load cell placed between the spring and the steel cube. We measure the dynamic force signals of the propagating waves with custom-made force sensors consisting of a piezo-electric disk embedded inside two halves of a stainless steel particle with radius $R=9.53$~mm. The sensor-particles are carefully constructed to resemble the mass, shape, and contact properties of the other spherical particles composing the rest of the crystal~\cite{NesterenkoSensor,dar05,dar05b,dar06,Job2005}.

\section{Theoretical model} We consider the 1D inhomogeneous crystal of $N$ beads as a chain of nonlinear oscillators~\cite{nesterenko1}:
\begin{equation}
\begin{aligned}
	m_{n}\ddot{u}_n&=A_{n}[\Delta_{n}+u_{n-1} - u_{n}]_{+}^{p} \\
	&- A_{n+1}[\Delta_{n+1}+u_{n} - u_{n+1}]_{+}^{p}\,,
\end{aligned}
\label{model}
\end{equation}
where $[Y]_+$ denotes the positive part of $Y$ (which signifies that adjacent particles interact only when they are in contact), $u_n$ is the displacement of the $n$th sphere (where $n \in \{1,\cdots,N\}$) around the static equilibrium, $m_n$ is the mass of the $n$th particle, and the coefficients $A_{n}$ depend on the exponent $p$ and the geometry/material properties of adjacent beads. The exponent $p=3/2$ represents the Hertz law potential between adjacent spheres~\cite{Johnson}. 
In this case, $A_{n}=\frac{2E}{3(1-\nu^2)}\left(\frac{R_{n-1}R_{n}}{R_{n-1} + R_{n}}\right)^{1/2}$, and the static displacement obtained from a static load $F_0$ is $\Delta_{n}=(F_0/A_{n})^{2/3}$ ~\cite{Johnson,nesterenko1}, where $R_n$ is the radius of the $n$th particle.

In order to study the linear spectrum of the inhomogeneous granular crystal, 
we linearize Eq. (\ref{model}) about the equilibrium state under the presence of the static load. This yields the following linear system~\cite{Herbold,Theo2009,our10}:
\begin{equation}
	m_{n}\ddot{u}_n=K_{n}(u_{n-1} - u_{n}) - K_{n+1}(u_{n} - u_{n+1})\,,
\label{linearEOM}
\end{equation}
where $K_n=\frac{3}{2}A_{n}^{2/3}F_{0}^{1/3}$. Following \cite{Theo2009}, we simplify Eq.(\ref{linearEOM}) to the eigensystem:
\begin{equation} \label{eigenvalueproblem}
	-\omega^{2}\mathbf{M}\mathbf{u}=\mathbf{\Lambda}\mathbf{u},
\end{equation}
where $\mathbf{M}$ is a $N\times N$ diagonal matrix with elements $M_{nn}=m_{n}$, and $\mathbf{u}$ is the displacement vector.
$\mathbf{\Lambda}$ is a $N\times N$ triagonal matrix with elements $\Lambda_{mn}=-[K_{n}+(1-\delta_{nN})K_{n+1}]\delta_{mn}+K_{n+1}(\delta_{mn-1}+\delta_{mn+1})$,
where $\delta$ is the Kronecker delta and we consider left-fixed and right-free boundary conditions. The right-free boundary assumption derives from the low stiffness of the static compression spring (Fig.~\ref{setup_spectrum}(a)) as compared to the stiffness of the particles in contact. 


\section{Single-defect: near-linear regime}
\begin{figure}[h]
\includegraphics[width=8.8cm,height=6.35cm,angle=0,clip]{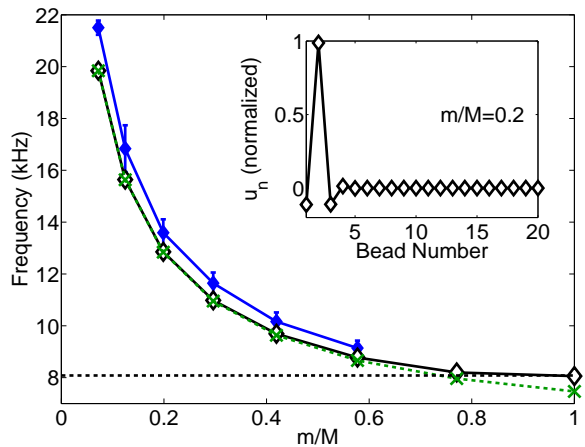}
\caption{[Color online] Frequency of the defect mode, with defect-bead placed at $n_{def}=2$, as a function of mass ratio $m/M$. Solid blue line (dark-grey, closed diamonds) corresponds to experiments, solid black line (open diamonds) to numerically obtained eigenfrequencies (see Eq. (\ref{eigenvalueproblem})), and green dashed line (light-grey, x markers) to the analytical prediction of the three-beads approximation (see Eq. (\ref{3bead})). The error bars account for statistical errors on the measured frequencies and are $\pm 2 \sigma$. Inset: The normalized defect mode for  $\frac{m}{M}=0.2$.}
\label{Fig2}
\end{figure}

In this section, we study 1D granular crystals that are homogeneous except for one light-mass defect bead at site $n_{def}$, as shown in Fig.~\ref{setup_spectrum}(a). Solving the eigenvalue problem of Eq.~(\ref{eigenvalueproblem}), for such a granular crystal, we obtain
the eigenfrequencies and the corresponding spatial profile of the modes of 
the system. The presence of the single light-mass defect generates a localized mode (see also \cite{Job,Theo2009}), 
centered at the defect site, which we will refer to as the defect mode. The defect mode amplitude decays exponentially away from the defect site and its frequency $f_d$ is such that $f_d > f_c$, where $f_c=\frac{1}{2\pi}\sqrt{\frac{4K_{RR}}{M}}$ is the upper cutoff frequency of the acoustic band of the homogeneous host crystal 
(where $K_{RR}=\frac{3}{2}A_{RR}^{2/3}F_{0}^{1/3}$ is the linear stiffness of the contact between two beads with radius $R$). The spatial profile of this mode consists of adjacent particles oscillating out of phase (see inset in Fig. (\ref{Fig2})). As the radius of the defect bead becomes smaller, the difference between $f_d$ and $f_c$ becomes larger, while the defect mode becomes more spatially localized. We observe that for the granular crystals studied here, with radii ratios of $\frac{r}{R}<0.7$, the defect mode involves the motion of up to approximately three beads, i.e., the 
displacements of the beads at $n\geq n_{def}+2$ and  $n\leq n_{def}-2$ 
are negligible. Because in this range of radii ratio the motion of the particles can be 
accurately approximated by three beads, we consider the particles at $n=n_{def}\pm 2$ as fixed walls, in order to find an analytical approximation for the frequency of the defect mode. Solving for the eigenfrequencies of this reduced three-bead system, we find that the mode corresponding to the out of phase motion can be analytically approximated by Eq. (\ref{3bead}) 
\begin{widetext} 
\begin{equation} \label{3bead}
	f_{3bead}=\frac{1}{2\pi}\sqrt{\frac{2K_{Rr}M+K_{RR}m+K_{Rr}m+ \sqrt{-8K_{Rr}K_{RR}mM+[2K_{Rr}M+(K_{RR}+K_{Rr})m]^{2}}}{2mM}}
\end{equation}
\end{widetext} 
%
where $K_{Rr}=\frac{3}{2}A_{Rr}^{2/3}F_{0}^{1/3}$ is the linear stiffness of the contact between a defect-bead and a bead of radius $R$.

We conduct experiments to identify the frequency of the 
defect mode in granular crystals with a single light-mass 
defect as shown in Fig.~\ref{setup_spectrum}(a). We place the defect particle at site $n_{def}=2$ (close to the actuator) so that the energy applied by the actuator, at the defect mode frequency, will not be completely attenuated by the uniform crystal, which acts as a mechanical frequency filter, before it arrives at the defect site. Because of the localized nature of the defect mode, placing a defect particle (of radius $r\le7.14$~mm) at site $n_{def}=2$ or further into the chain makes nearly no difference on the frequency of the defect mode. For instance, for a defect particle of radius $r=7.14$~mm, we numerically calculate (using Eq.~\ref{eigenvalueproblem}) the difference in the defect mode frequency for the cases where a defect particle is placed at site $n_{def}=2$ or $n_{def}=10$, to be $3$~Hz. Conversely, because of the presence of the fixed boundary and the larger localization length of the defect mode, for a defect particle of $r=8.73$~mm, we calculate the difference in defect mode frequency, between sites $n_{def}=2$ and $n_{def}=10$, to be $68$~Hz. The defect particles are stainless steel spheres of smaller radii, $r=[3.97,~4.76,~5.56,~6.35,~7.14,~7.94]$~mm, and 
measured masses of $m=[2.08,~3.60,~5.73,~8.54,~12.09,~16.65]$~g, respectively. We experimentally characterize the linear spectrum of this system by applying low amplitude (approximately $200$~mN) bandwidth limited noise ($3-25$~kHz for the two smallest defect particles, and $3-15$~kHz otherwise) via the piezoelectric actuator. We calculate the transfer functions, specific to the sensor location, by averaging the Power Spectral Densities (PSD~\cite{PSD}) of 16 force-time histories, measured with the embedded sensors, and dividing by the average PSD level in the $3-8$~kHz range (corresponding to the acoustic band). We embed sensors in particles at sites $n=4$ and $n=20$. In Fig.~\ref{setup_spectrum}(b) we show the transfer functions for the granular crystal with defect radius $r=5.56$~mm. The red (light-grey) and blue (dark-grey) curves are the transfer functions for the sensors at sites $n=4$ and $n=20$, respectively. We denote the experimental cutoff frequency by the triangular marker (found by identifying the last peak in the acoustic band) and defect frequency as the diamond marker on the $n=4$ transfer function. The vertical lines denote the theoretically determined upper cutoff frequency of the acoustic band and the defect frequency (Eq.~\ref{3bead}). The presence of the defect mode can be clearly
identified in the vicinity of the defect (at $n=4$), but is not visible far from the defect (at $n=20$). 

We repeat the process of measuring the transfer function and identifying the defect mode frequency 16 times, re-assembling the crystal after each repetition. In Fig.~\ref{Fig2}, we plot the average frequency of the 16 experimentally identified defect 
modes as a function of the mass ratio $\frac{m}{M}$ (blue [dark-grey] solid line
connecting the closed diamonds). We also plot, for comparison, the defect frequency predicted by the analytical expression of Eq.~(\ref{3bead}) (green [light-grey] dashed line connecting the crosses), and the
numerical eigenanalysis of Eq. (\ref{eigenvalueproblem}) corresponding to the 
experimental setup (black solid line connecting
the open diamonds). The error bars on the experimental data are $\pm 2 \sigma$ 
where $\sigma$ is the standard deviation of the identified defect frequencies over the 16 repetitions. Comparing the 
analytical three-bead approximation with the numerical
eigenfrequencies, we find an excellent agreement for mass ratios 
of $\frac{m}{M}<0.6$. 
Comparing the experimental data with the numerics, we find an upshift of $5-10$\%, similar to the upshift observed in \cite{our10,our11}. For the $r=5.56$~mm defect, the average experimental defect frequency is $f_d^{exp}=13.59$~kHz and the average experimental cutoff frequency is $f_c^{exp}=8.36$~kHz. In comparison, the theoretical 3-bead approximation gives a defect frequency of $f_d^{3bead}=12.84$~kHz and the eigenproblem of Eq. (\ref{eigenvalueproblem}) gives a defect frequency of $f_d^{num}=12.85$~kHz, while the analytically calculated
cutoff frequency was $f_c=8.02$~kHz.

Possible reasons for these upshifts have been identified 
in \cite{our10, our11} and the references therein, such as error in the material parameters, nonlinear elasticity, surface roughness, dissipative mechanisms and misalignment of the particles. We note that a systematic error in the measurement of the static load could also cause such an upshift. Nevertheless, it is clear from Fig.~\ref{Fig2} that the functional dependence of the relevant frequencies on the mass ratio (of defect to regular beads) is accurately captured by our analytical and numerical results.

%


\section{Two-defects: near-linear regime} We study granular crystal configurations with two identical light-mass 
defects to better understand the effects of increasing heterogeneity on the spectral response of the system. The localized mode due to the presence of a light-mass defect (for mass ratios $\frac{m}{M}<0.6$) has a spatial localization length of about three particles (larger particles have a greater localization length, and smaller 
particles have a shorter localization length), as described in the one-defect case and shown in Fig.~\ref{Fig2}. We can thus expect that two light-mass defects placed far from each other in a granular crystal (sufficiently outside this localization length) would have similar frequencies and mode shapes independent of the presence of the other. However, as the two defect particles are brought closer together (within the localization length), each mode influences the other. For a sufficiently small mass ratio, this results in the creation of two defect modes at different frequencies; one with the defect particles moving out of phase, and the other with the defect particles moving in phase. For the case of nearest-neighbor identical defects, our theoretical analysis can be extended by using a four-particle analogy. In this case, using the notation
$s_1=K_{Rr} (M+m) + K_{RR} m$, $s_2=-4 K_{Rr} K_{RR} M m +  
    (K_{RR} m + K_{Rr} (M + m))^2$,
$s_3=s_1+ 2 K_{rr} M$ and $s_4=-4 (2 K_{rr} K_{RR} + K_{Rr} (2 K_{rr} + K_{RR})) M m + (2 K_{rr} M + K_{RR} m +
     K_{Rr} (M + m))^2$, we obtain the following frequencies
\begin{eqnarray}
f_{4bead}^{(1)} &=&\frac{1}{2\pi} \sqrt{\frac{1}{2 M m} \left( s_1 \pm \sqrt{s_2} \right)},
\label{4b1}
\\
f_{4bead}^{(2)} &=&\frac{1}{2\pi} \sqrt{\frac{1}{2 M m} \left( s_3 \pm \sqrt{s_4} \right)}.
\label{4b2}
\end{eqnarray}
The two highest frequencies correspond to the linear defect mode frequencies.
Naturally, this analytical approach can be extended to more distant
defects, although we do not present such algebraically intensive
cases here.

\begin{figure}[h]
\begin{center}
\includegraphics[width=8.3cm,height=3.66cm]{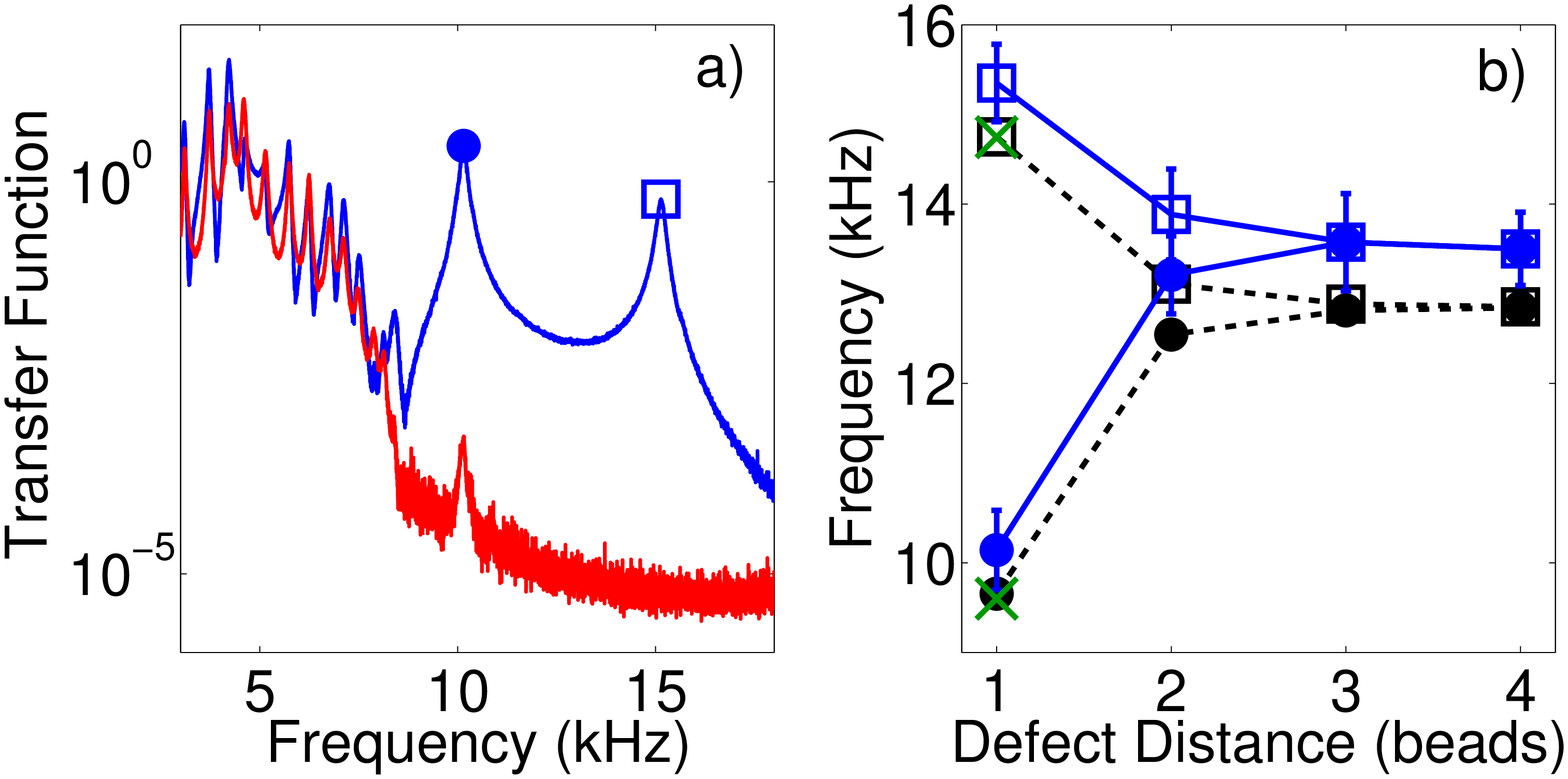}
\includegraphics[width=8.3cm,height=2.34cm]{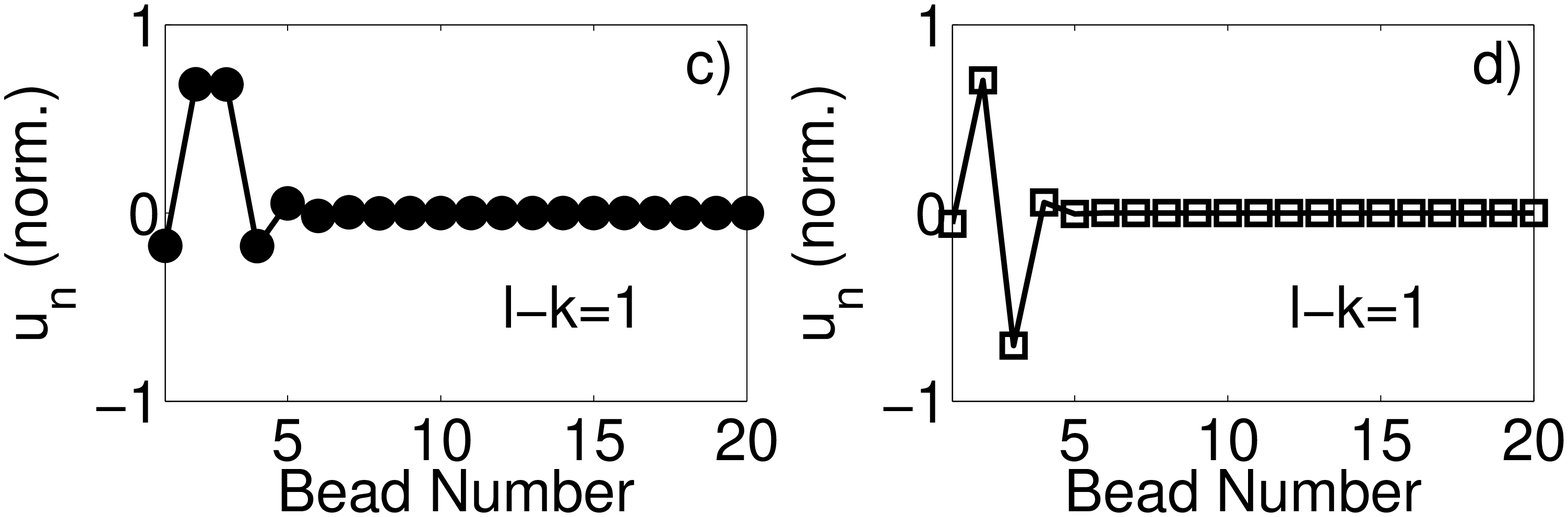}
\end{center}
\caption{\label{twodefects}[Color online] (a) Experimental transfer functions for a granular crystal with two defect-beads of mass ratio $\frac{m}{M}=0.2$ at $n_{def}=2$ and $n_{def}=3$ (in contact). Blue (dark-grey) [red (light-grey)] curve corresponds to transfer function obtained from the force signal of a custom sensor placed at $n=4$ [$n=20$]. (b) Frequencies of the defect modes as a function of the distance between them. The solid line denotes experimental data, the dashed line the numerically obtained eigenfrequncies, and the x markers the frequencies from the analytical expresssions of Eqs.~(\ref{4b1})-~(\ref{4b2}). (c),(d) The normalized defect mode shapes corresponding to the defect modes identified in (a) with frequency of the same marker-type.}
\end{figure}

In Fig.~\ref{twodefects}, we show the behavior of two $r=5.56$~mm defects in a $N=20$ particle granular crystal under $F_0=20$~N static load (similiar to the configuration shown in Fig.~\ref{setup_spectrum}(a)), where the first defect is at site $k=n_{def1}=2$ and the second defect is at a variable position between site $l=n_{def2}=3$ and $l=n_{def2}=6$. We use the same experimental method as in the single defect case except now we use a noise range between $3-20$~kHz and we place the first sensor at $n=l+1$. We show the experimentally determined PSD transfer function for the case of  $l-k=1$ in Fig.~\ref{twodefects}(a), with sensors at site $n=4$ (blue [dark-grey]) and $n=20$ (red [light-grey]). As described in ~\cite{Theo2009}, the existence of two separate defect modes for the case where the defect particles are adjacent to each other ($l-k=1$), depends on the mass ratio of the defect particles to those of the rest of the crystal. Here the mass ratio is such that two modes are present, as can be seen in the blue (dark-grey) curve in Fig.~\ref{twodefects}(a). The two distinct modes, which we denote by the open square and closed circular markers, have frequencies above the acoustic band. The square markers denote the mode with defect beads moving out of phase, and the closed circular marker corresponds to the mode with defect particles moving in phase, as shown by the 
numerically calculated eigenmodes in Fig.~\ref{twodefects}(c) and (d) respectively~\cite{Theo2009}. 
In Fig.~\ref{twodefects}(b) we plot the experimentally determined frequencies 
of both modes as a function of the inter-defect particle distance ($l-k$). The 
solid blue (dark-grey) lines are the experimental data, and the dashed black lines are the 
frequencies obtained from solving the eigenvalue problem of 
Eq.~(\ref{eigenvalueproblem}). The green (light-grey) x-markers denote the frequencies calculated with Eqs.~(\ref{4b1})-(\ref{4b2}),
for the $l-k=1$ case. It is evident that the analytical results agree closely with the numerically calculated eigen-frequencies. The error bars on the experimental data correspond to the $\pm 2 \sigma$ standard deviation as calculated in the single-defect case. We see close qualitative agreement between the experimental data and the numerical predictions, but also the same systematic upshift as 
observed in the single defect case and~\cite{our10,our11}. From Fig.~\ref{twodefects}(b) we can see that as the defects are placed three or more particles apart, the frequencies of the defect modes converged to approximately the same value, suggesting the defects respond independently of each other. 

\section{Single defect: nonlinear localized modes} As shown in \cite{Theo2009}, the interplay of the inherent nonlinearity
of the granular crystal with the linear localization due to the defect
results in the presence of robust nonlinear localized modes (NLMs). The frequency of these modes depends not only on the 
static load and the material values of the beads but also on the amplitude of the oscillations.
In order to find this dependence, we apply Netwon's method (see \cite{Theo2009}
and references therein) for the experimental, single-defect, configuration of Fig. 1(a). For the numerical calculations in this section, we calculate experimental contact coefficients following a procedure similar to the one described in \cite{our10}. The experimental contact coefficients obtained are $A_{RR}^{exp}=10.79$~N/$\mu m^{3/2}$ for the contact between two $R=9.53$~mm beads and $A_{rR}^{exp}=9.95$~N/ $ \mu m^{3/2}$ for the contact between the $R=9.53$~mm and the $r=5.56$~mm beads. In comparison, the values of the coefficient $A$, as calculated by the material values and used for the previous sections of the paper, are $A_{RR}=9.76$~N/$\mu m^{3/2}$ and $A_{Rr}=8.38$~N/ $ \mu m^{3/2}$. 

In Fig. 4(a), we show the 
frequency of the numerically determined NLM as a function of the averaged dynamic force for the particle at site $n=3$. The latter corresponds to the average of the two dynamic contact forces adjacent to the particle,
which is analogous to what is measured experimentally by the dynamic force sensor~\cite{dar05}. In Fig. 4(b), we plot the numerically determined normalized
NLM shape at $f_b=13.28$~kHz. Comparing this NLM shape to the linear mode shape of the same frequency (see inset of Fig.~\ref{Fig2}), we can see that the NLM has a slightly modified (more asymmetric) spatial profile.

The experimental setup used for the study of the NLMs is the same as in the case of the linear single defect experiments (as shown in Fig.~\ref{setup_spectrum}(a)) except we place sensors in particles at sites $n=3$, $n=5$, and $n=20$. 
Additionally, we replace the $n=1$ particle with an embedded actuator particle, so as to apply high amplitude (approximately $10$~N), short time pulse (approximately $100$~$\mu$s) perturbations directly to the defect particle. Exciting such a pulse creates an initial condition in the crystal that resembles the predicted defect NLM shape. The embedded actuator particle is similar in construction to the sensors but with a piezoceramic construction/geometry more appropriate for high force amplitude actuation (Piezomechanik PCh 150/5x5/2 Piezo-chip). 

\begin{figure}[h]
\begin{center}
\includegraphics[width=8.8cm,height=6.35cm,angle=0]{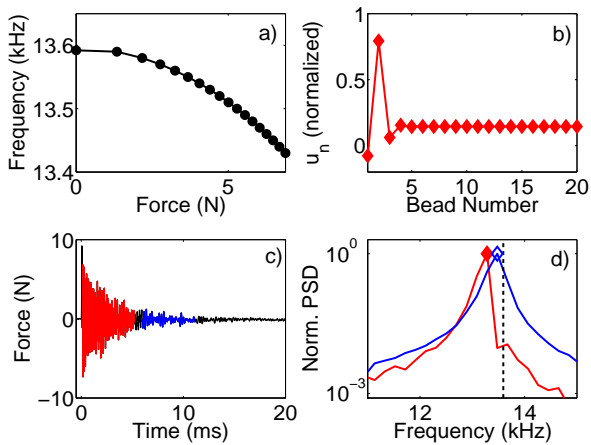}
\caption{\label{Fig4}[Color online] (a) Numerical frequency continuation of the nonlinear defect modes corresponding to the experimental setup in Fig.~\ref{setup_spectrum}(a). (b) Numerically calculated spatial profile of the nonlinear localized mode with frequency $f_{def}=13.28$~kHz. (c) Measured force-time history of a sensor at site $n=3$, where a high amplitude, short width, force pulse is applied to the granular crystal. (d) Normalized PSD for the measured time regions of the same color in (c); closed and open diamonds correspond to the high and low amplitude time regions respectively. The vertical dashed line is the mean experimentally determined linear defect mode frequency.}
\end{center} 
\end{figure}

The force-time history of the dynamic force measured by the sensor at site $n=3$ is shown in Fig.~\ref{Fig4}(c). A sharp excitation is evident at time $t=0$, followed by periodic oscillations with a decaying envelope, due to the inherent dissipation in the system. As shown by the parametric continuation in Fig.~\ref{Fig4}(a), NLMs corresponding to the defect mode at higher amplitudes have a frequency deeper into the gap than its linear counterpart. However, for the amplitudes observed here this is only a slight shift (up to $200$~Hz over $7$~N). 

We study, in more detail, the response of two selected time regions of the force-time history shown in Fig.~\ref{Fig4}(c), to experimentally demonstrate the frequency shift characteristic of higher amplitude NLMs. The two non-overlapping time regions are of length $T=5.1$~ms. The red (light-grey) time region begins immediately following the arrival of the initial actuated pulse, and presents a maximum amplitude of $7$~N. The blue (dark-grey) time region starts $T=6$~ms after the begining of the previous time region, and presents a maximum amplitude of $1.3$~N. We calculate PSDs for both time regions (frequency resolution $\delta f=195$~Hz) as shown in Fig.~\ref{Fig4}(d). The PSDs shown in Fig.~\ref{Fig4}(d) correspond to the time regions of the same color shown in Fig.~\ref{Fig4}(c). Here, the PSDs are normalized by dividing the PSD by the peak PSD amplitude of the identified defect mode. It is evident that the peak in the PSD spectrum corresponding to the time region with larger force amplitude presents a lower characteristic frequency (i.e., it is further into the gap) with respect to the peak representing the time region with lower force amplitudes. This is in agreement with the shift predicted by the parametric continuation analysis shown in Fig.~\ref{Fig4}(a). The peak frequency of the PSD of the high force amplitude time region is $f_{def}=13.28$~kHz, and the peak frequency of the PSD of the low amplitude time region is $f_{def}=13.48$~kHz, where $f_{def}=13.48$~kHz is closer to the mean experimentally determined linear defect mode frequency (shown by the dashed line in Fig.~\ref{Fig4}(d)). 

\section{Conclusions} We studied the response of statically compressed granular crystals containing light-mass defects, and characterized their near-linear spectra by applying continuous excitation. We demonstrated that such chains support localized modes with frequencies above that of their acoustic band cutoff, using approximate few-bead analytical calculations, numerics, and experiments. The number of supported localized modes depends on the number of defects, while their frequencies depend on the inter-defect distance, on the ratio $\frac{m}{M}$ of defect to regular masses (and the geometric/elastic properties of the beads), and on the static load.  We also briefly described the nonlinear generalizations of such modes, departing from the near-linear regime, and showed a downshift of the corresponding defect mode frequencies with increasing amplitude. This study is important for understanding the interplay of disorder and nonlinearity in discrete systems, and the results reported may be relevant in the design of applications involving vibrational energy trapping.


\acknowledgments 
We thank St\'{e}phane Job for help with the experimental setup. GT and PGK acknowledge support from the
A.S. Onassis Public Benefit Foundation through RZG 003/2010-2011 and PGK also through NSF-CMMI-1000337. 
CD acknowledges support from NSF-CMMI-969541 and NSF-CMMI-844540 (CAREER).

\end{document}